\documentclass[sigconf,nonacm]{acmart}
\usepackage{multirow}
\usepackage{pifont}

\newcommand{\cmark}{\ding{51}}
\newcommand{\xmark}{\ding{55}}
\AtBeginDocument{%¡
  }

\author{Kunlin Wu$^1$, Yanning Wang$^1$, Haofeng Tan$^2$, Boyi Chen$^1$, Teng Fei$^3$,\\ Xianping Ma$^4$, Yang Yue$^1$, Zan Zhou$^5$, and Xiaofeng Liu$^1$ \\[15pt]}

\affiliation{
  \institution{
    $^1$The Hong Kong University of Science and Technology (Guangzhou), China \quad
    $^2$University of South Carolina, USA \\
    $^3$University of Canterbury, New Zealand \quad
    $^4$Southwest Jiaotong University, China \quad \\
    $^5$Beijing University of Posts and Telecommunications, China
  }
  \country{} % 留空以节省最后一行空间
}
\begin{document}

\title{Geo2Sound: A Scalable Geo-Aligned Framework for Soundscape Generation from Satellite Imagery}

\begin{abstract}
Recent image-to-audio models have shown impressive performance on object-centric visual scenes. However, their application to satellite imagery remains limited by the complex, wide-area semantic ambiguity of top-down views. While satellite imagery provides a uniquely scalable source for global soundscape generation, matching these views to real acoustic environments with unique spatial structures is inherently difficult. To address this challenge, we introduce Geo2Sound, a novel task and framework for generating geographically realistic soundscapes from satellite imagery. Specifically, Geo2Sound combines structural geospatial attributes modeling, semantic hypothesis expansion, and geo-acoustic alignment in a unified framework. A lightweight classifier summarizes overhead scenes into compact geographic attributes, multiple sound-oriented semantic hypotheses are used to generate diverse acoustically plausible candidates, and a geo-acoustic alignment module projects geographic attributes into the acoustic embedding space and identifies the candidate most consistent with the candidate sets. Moreover, we establish SatSound-Bench, the first benchmark comprising over 20k high-quality paired satellite images, text descriptions, and real-world audio recordings, collected from the field across more than 10 countries and complemented by three public datasets. Experiments show that Geo2Sound achieves a SOTA FAD of 1.765, outperforming the strongest baseline by 50.0\%. Human evaluations further confirm substantial gains in both realism (26.5\%) and semantic alignment, validating our high-fidelity synthesis on scale. Project page and source code: \url{https://github.com/Blanketzzz/Geo2Sound}

\end{abstract}

%\begin{CCSXML}
%<ccs2012>
%  <concept_id>10010405.10010432.10010437</concept_id>
%  <concept_desc>Applied computing~Earth and atmospheric sciences</concept_desc>
%  <concept_significance>500</concept_significance>
%  <concept_id>10010147.10010371.10010382</concept_id>
%  <concept_desc>Computing methodologies~Multimedia and multimodal processing</concept_desc>
%  <concept_significance>500</concept_significance>
%</ccs2012>
%\end{CCSXML}

%\ccsdesc[500]{Computing methodologies~Scene understanding}

%\keywords{Cross-modal Soundscape Generation, Satellite Imagery, Multimodal Geospatial Alignment}


\begin{teaserfigure}
  \captionsetup{skip=1pt}
  \includegraphics[width=\textwidth]{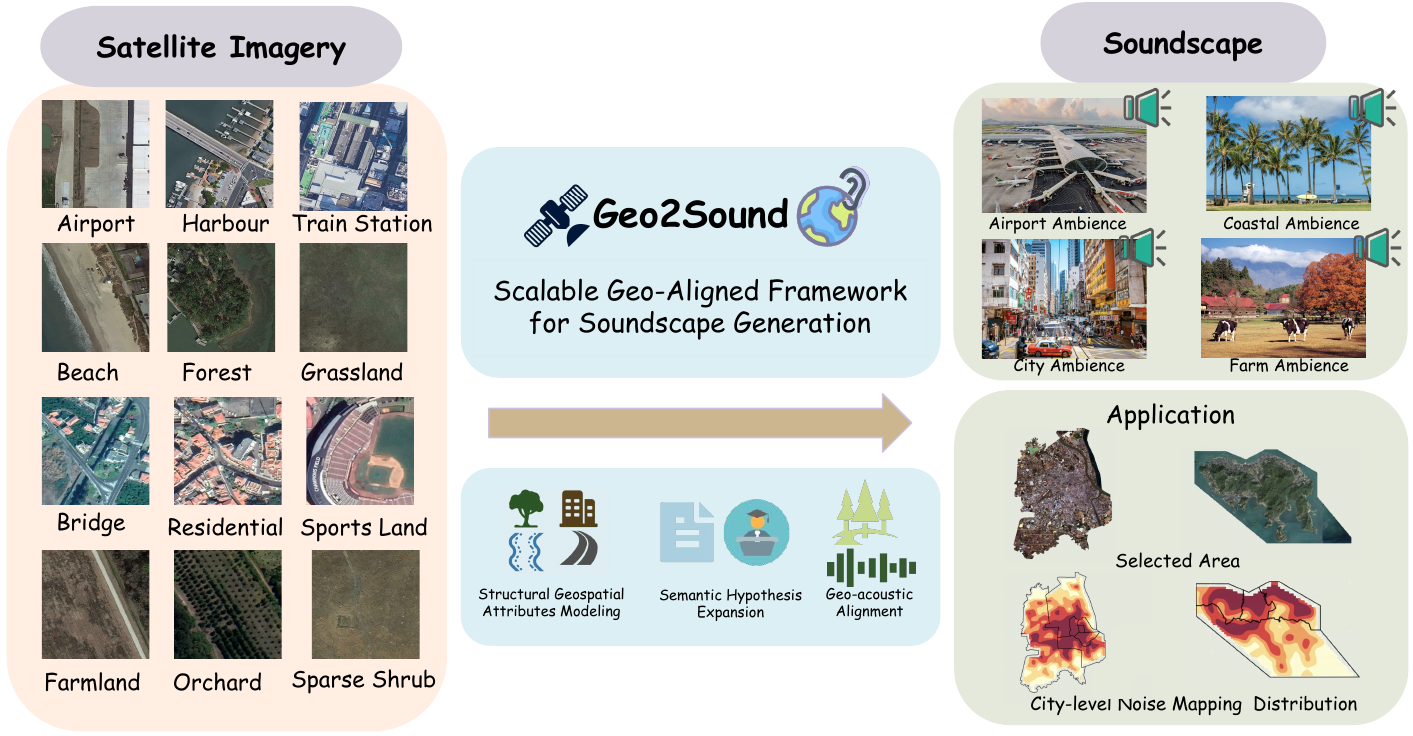}
  \caption{Overview of Geo2Sound. Given satellite imagery, Geo2Sound generates geographically plausible soundscapes through structural geospatial attributes modeling, semantic hypothesis expansion, and geo-acoustic alignment, and further supports downstream applications such as city-level noise mapping.}
  \label{fig:teaser}
\end{teaserfigure}

\maketitle
\pagestyle{plain}

\section{Introduction}

Soundscapes are fundamental to daily experience, shaping how people perceive and interact with their surroundings \cite{Schafer1977, Truax2001, Pijanowski2011}. They provide essential information for noise monitoring and urban planning \cite{soundscape1,soundscape2,soundscape3,soundscape4,soundscape5}, and are increasingly important for multimedia applications such as digital twins and interactive maps, which require realistic audio that matches geographic context and spatial structure \cite{Dagli2024See2Sound, Wang2024TiVA, rong2025audiogenie, Wang2024BridgeALM}. For instance, the steady hum of engines often signals an industrial area, while distant crowd chatter conveys the lively atmosphere of a busy market. However, obtaining geographically consistent soundscapes over large areas remains challenging, since existing research relies heavily on field recordings, which are costly and difficult to collect at scale, particularly in remote, underdeveloped, or rapidly changing regions \cite{Brown2011, Lionello2020, Chasmai2024iNatSounds, freesound}. In contrast, high-resolution satellite imagery is globally and continuously available, providing rich spatial context about land use, infrastructure, and environmental conditions \cite{Zhu2022LandChange, Li2024UrbanLandUseReview, Li2022LowToHighLCM, Chen2016InformationFromImagery}. This setting motivates the task of satellite-to-soundscape generation, illustrated in Figure~\ref{fig:teaser}.

Recent image-to-audio (I2A) models \cite{vision2audio5, vision2audio2, vision2audio3, vision2audio4, vision2audio6} and multimodal-to-audio (M2A) systems \cite{wu2024nextgpt, tang2023codi2, rong2025audiogenie, Tian2025AudioX} have shown promising performance in generating audio from visual inputs such as street-view scenes and object-centric images. Some recent efforts have also explored satellite imagery for soundscape retrieval and mapping \cite{khanal2023learning, khanal2025sat2sound}. However, direct soundscape generation from satellite imagery remains underexplored and faces three key challenges. First, overhead imagery lacks \textbf{structured geospatial semantics} for acoustic reasoning. Existing audio generation frameworks are primarily developed for ground-level scenes in which dominant sound sources, such as vehicles or animals, are visually explicit \cite{vision2audio6, audio_dataset3, Zhao2023, Wang2025SounDiT}. In contrast, satellite imagery compresses diverse environmental elements into a top-down spatial layout, making it difficult to infer acoustically meaningful scene descriptors directly from raw visual patterns \cite{remote4, Jakubik2024Prithvi, Chen2025}. Second, static satellite imagery introduces \textbf{one-to-many acoustic ambiguity.} Unlike ground-level observations that benefit from local appearance and temporal motion cues \cite{vision2audio5, vision2audio2}, top-down views often correspond to multiple plausible soundscapes \cite{Heidler2022SoundingEarth, Chen2025}. For example, an industrial rooftop may indicate either an actively operating plant or a quiet warehouse. Without diverse acoustic semantic interpretations, general-purpose models tend to produce generic or mismatched audio. Third, soundscape generation requires \textbf{broader geospatial context.} Soundscapes are shaped not only by local appearance but also by surrounding land use, road structure, density, and human activity \cite{urban3, soundscape4}. As a result, visually similar local regions may correspond to very different acoustic environments depending on their larger geographic setting \cite{soundscape3, Huang2024, remote4}. Without modeling such broader context, it remains difficult to generate geographically plausible soundscapes at scale \cite{Jakubik2024Prithvi, remote4}.

To address these challenges, we propose \textbf{Geo2Sound}, a scalable geo-aligned framework for soundscape generation from satellite imagery. Geo2Sound consists of three components: (1) structural geospatial attributes modeling, which summarizes acoustically relevant environmental cues from overhead scenes; (2) semantic hypothesis expansion, which generates multiple plausible soundscape interpretations for the same spatial scene; and (3) geo-acoustic alignment, which selects the candidate most consistent with the surrounding geographic environment. To support this task, we further establish SatSound-Bench, the first large-scale benchmark for satellite-to-soundscape generation, comprising paired satellite imagery, textual descriptions, and audio data collected from both real-world field recordings and complementary public datasets.

The main contributions of this work are as follows:
\begin{itemize}
    \item \textbf{A new task and benchmark.} We formulate satellite imagery-to-soundscape generation as a new multimodal generative task and establish SatSound-Bench, the first large-scale benchmark comprising paired satellite imagery, text, and audio for this setting. The benchmark integrates both real-world field recordings and complementary public datasets to support evaluation across diverse environments.

    \item \textbf{A scalable geo-aligned framework.} We propose Geo2Sound, which addresses the unique challenges of overhead imagery through structural geospatial attributes modeling, semantic hypothesis expansion, and geo-acoustic alignment. This design explicitly connects scene structure, semantic ambiguity, and geographic context in a unified generation pipeline.

    \item \textbf{State-of-the-art performance.} Extensive experiments and human evaluations show that Geo2Sound consistently outperforms existing I2A and M2A baselines in both acoustic quality and geographic plausibility. The results further demonstrate the value of combining semantic diversity with geographic consistency for soundscape generation from satellite imagery.
\end{itemize}

\section{Related Work}

\begin{figure*}[t]
  \centering
  \includegraphics[width=\textwidth]{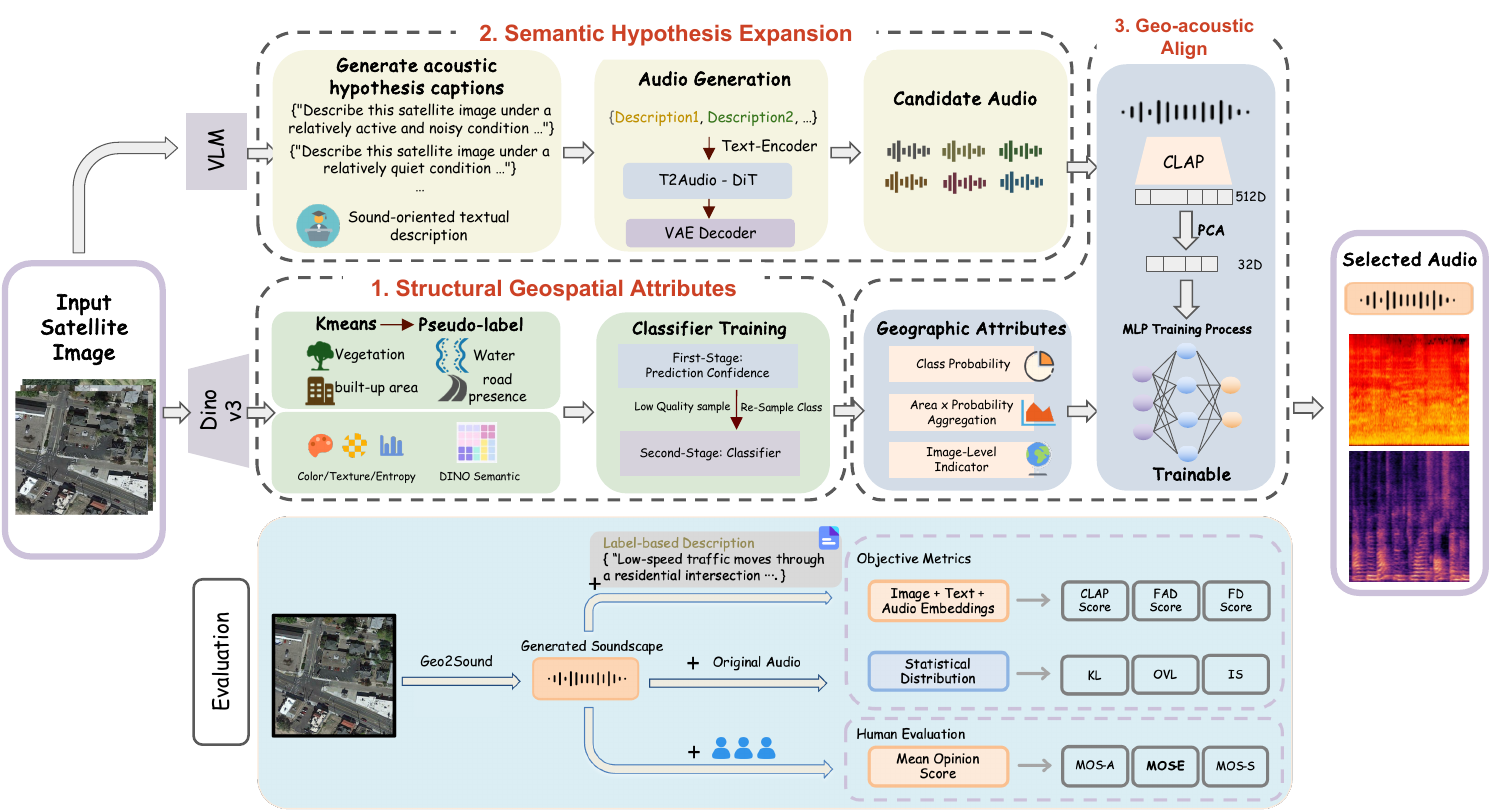}
  \caption{Overview of Geo2Sound. The framework combines structural geospatial attributes, semantic hypothesis expansion, and geo-acoustic alignment to generate geographically plausible soundscapes from satellite imagery, followed by objective and human evaluation.}
  \label{fig:pipeline}
\end{figure*}

\subsection{Cross-Modal Audio Generation}

Recent studies in cross-modal audio generation have explored conditioning audio synthesis on diverse input modalities. Among them, image-to-audio (I2A) generation aims to produce plausible soundscapes from visual observations \cite{Li2025SoundingObject, vision2audio6, vision2audio2, vision2audio5, vision2audio3, Wang2024TiVA, Chen2025MultiFoley}. Building upon this paradigm, multimodal-to-audio (M2A) frameworks \cite{rong2025audiogenie, tang2023codi2, Tian2025AudioX} further integrate multiple inputs to improve synthesis through cross-modal fusion. However, these models remain difficult to apply to satellite imagery, where soundscapes are shaped by distributed spatial elements rather than isolated sound-producing entities. In addition, static overhead scenes often correspond to multiple plausible acoustic states \cite{Jakubik2024Prithvi, khanal2023learning, khanal2025sat2sound}.

At the same time, the strong semantic understanding capability of recent vision-language models (VLMs) \cite{Radford2021CLIP, Alayrac2022Flamingo, Li2023BLIP2, Dai2023InstructBLIP, Li2019VisualBERT} and the generative capability of text-to-audio (T2A) models \cite{text2audio1, text2audio2, text2audio3, text2audio4, text2audio5, text2audio6, text2audio7} suggest a promising scalable path toward satellite-to-soundscape generation. However, combining these capabilities remains difficult due to a critical data gap: large-scale paired satellite-image--text--audio data are still largely unavailable. Moreover, existing VLM and T2A frameworks are not designed to transform distributed geospatial structure into acoustically meaningful conditions, nor to ensure geo-acoustic alignment during soundscape generation, which may lead to acoustically plausible but geographically inconsistent soundscapes \cite{khanal2023learning, khanal2025sat2sound}. These limitations motivate our scalable geo-aligned framework and the construction of SatSound-Bench for satellite-to-soundscape generation research.

\subsection{Geospatial Semantics for Satellite Imagery Understanding}

Recent advances in satellite imagery foundation models have substantially improved the semantic understanding of overhead scenes \cite{Cong2022SatMAE, Jakubik2024Prithvi, liu2023remoteclip, wang2024skyscript, hu2023rsgpt, Shabbir2025GeoPixel}. These models enable increasingly rich modeling of geospatial context from satellite observations and support a variety of downstream tasks, including land-cover classification, urban analysis, visual question answering, and cross-modal retrieval \cite{Klemmer2023SatCLIP, kuckreja2023geochat, liu2023remoteclip, khanal2025sat2sound}.

However, existing geospatial semantics and alignment frameworks are not designed for soundscape generation. In particular, they do not transform implicit geospatial structure into acoustically meaningful generation conditions, nor do they explicitly model geo-acoustic alignment for direct soundscape synthesis. The closest satellite-soundscape works, including GeoCLAP \cite{khanal2023learning} and the later framework \cite{khanal2025sat2sound}, primarily address soundscape mapping or multimodal retrieval through shared embedding learning rather than direct soundscape generation.

Our work addresses this gap by extending geospatial semantics from visual understanding to direct soundscape generation. Specifically, satellite-derived geographic attributes are transformed into acoustically meaningful conditions and further constrained through geo-acoustic alignment during candidate selection. This design moves beyond representation learning and retrieval by explicitly coupling geospatial scene understanding with soundscape generation. In the next section, we describe how Geo2Sound operationalizes this idea through structural geospatial attributes modeling, semantic hypothesis expansion, and geo-acoustic alignment.

\section{Method}

The overall framework is illustrated in Figure~\ref{fig:pipeline}. Geo2Sound consists of three components: structural geospatial attributes modeling, semantic hypothesis expansion, and geo-acoustic alignment.

\begin{figure*}[t]
  \centering
  \includegraphics[width=\textwidth]{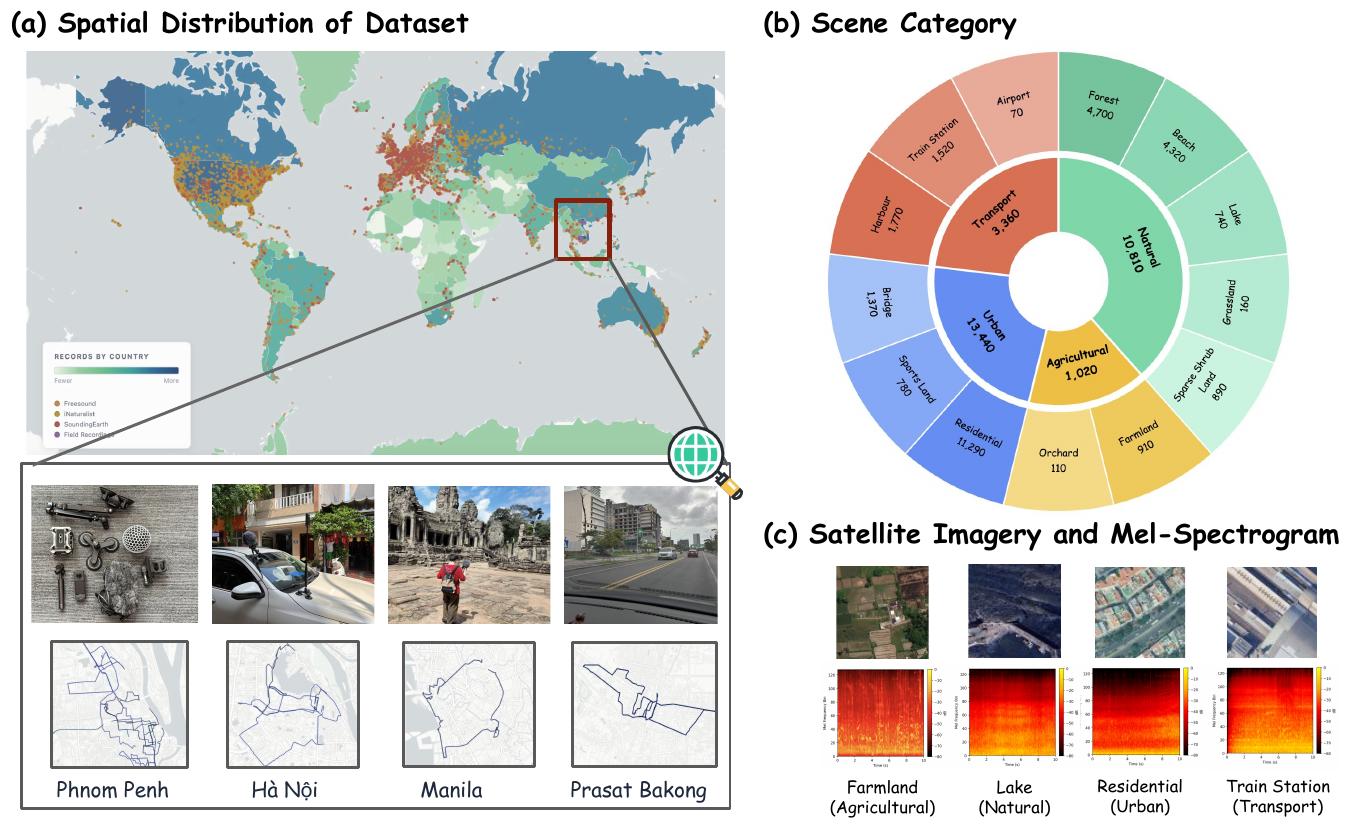}
  \caption{Overview of SatSound-Bench. (a) Global spatial distribution of audio recordings from both field-collected data and public datasets, with field recordings captured using vehicle-mounted systems across diverse environments. (b) Scene category composition, organized into four coarse domains---transport, urban, natural, and agricultural---and thirteen fine-grained scene categories. (c) Representative examples of paired satellite imagery and corresponding mel-spectrograms, illustrating the alignment between overhead visual structures and environmental sound patterns.}
  \label{fig:satsoundbench}
\end{figure*}

\subsection{Structural Geospatial Attributes Modeling}

To address the lack of structured geospatial semantics for acoustic reasoning in overhead imagery, we design a module that transforms raw overhead visual imagery into a set of structured geographic attributes without relying on manual annotations. We first extract dense patch-level embeddings from satellite imagery using a fixed visual backbone. To obtain a spatially coherent decomposition of the scene, we partition these patch embeddings into clusters using K-means and upsample the resulting label map to the original image resolution. For each cluster, we compute visual descriptors, including RGB/HSV statistics, grayscale texture cues, and edge density, and concatenate them with the cluster-level mean embedding.

We then perform pseudo-label generation by heuristic scoring over the same cluster-level appearance statistics, and use the resulting pseudo-labels to train a two-stage random forest classifier. In the first stage, the model estimates prediction confidence and filters out low-quality pseudo-labeled samples. In the second stage, the classifier is retrained on the resulting high-confidence subset and serves as the final cluster-level predictor. At inference, the classifier outputs class probabilities over multiple categories, such as vegetation, water, built-up areas, and roads.

These class probabilities are then aggregated across clusters using area-weighted accumulation, corresponding to the area $\times$ probability aggregation shown in Figure~\ref{fig:pipeline}. This produces image-level geographic attributes that summarize the structural composition of the scene. To further capture scene-level spatial diversity, we compute the Shannon entropy of the aggregated proportions. Together, the class proportions and this diversity term form a compact geographic descriptor, which is used as the structural input to the subsequent geo-acoustic alignment module. 

\subsection{Semantic Hypothesis Expansion}

To address one-to-many acoustic ambiguity, we introduce a semantic hypothesis expansion strategy that constructs a structured set of sound-oriented textual descriptions for each scene. Instead of relying on a single caption, we generate multiple semantic hypotheses that reflect variations in dominant sound sources, environmental context, and activity intensity. Each hypothesis is then used as conditioning input to a text-to-audio model, producing multiple candidate audio realizations. 

In our implementation, each input scene yields six candidate audio realizations derived from three prompt variants, corresponding to basic, relatively quiet, and relatively noisy acoustic conditions. By expanding the candidate sets, this design increases the chance of producing a soundscape that matches both the acoustic hypothesis and the spatial scene. This candidate-based design is particularly important for overhead imagery, where a single scene may correspond to multiple plausible acoustic states. It therefore preserves acoustic diversity before geographic context is introduced for final disambiguation.

Compared with single-caption conditioning, semantic hypothesis expansion improves both semantic alignment and distributional consistency of generated audio. More importantly, it explicitly separates candidate generation, allowing acoustic diversity to be explored before geo-acoustic alignment is imposed. This expanded candidate set provides the basis for the subsequent geo-acoustic alignment step. Rather than enforcing geographic constraints during generation, we apply them afterward to identify the candidate that is most compatible with the spatial context.

\subsection{Geo-acoustic Alignment Module}

To address the lack of broader geospatial context in candidate selection, we introduce a geo-acoustic alignment module that identifies the most geographically plausible output from the hypothesis-expanded candidates. The key idea is to project geographic indicators into an acoustic embedding space and measure their compatibility with candidate audio embeddings.

Specifically, we use the geographic attributes derived above to represent the spatial structure of the scene, and train a lightweight projection network on paired geographic--audio samples to map these attributes into the acoustic embedding space. This projected geographic representation serves as a query vector that captures the expected acoustic characteristics of the environment.

Candidate audio clips are encoded into the shared acoustic embedding space. We then compute cosine similarity between the projected geographic embedding and each candidate audio embedding in the shared acoustic space, and select the candidate with the highest compatibility score as the final output. This design enables geographic context to act as a scene-level acoustic prior over the candidate set. As a result, geographic plausibility is improved without restricting the diversity introduced during candidate generation.

By operating on the hypothesis-expanded candidate set, this module allows geographic context to act as a global acoustic prior over multiple plausible realizations of the same scene. In this way, the geo-acoustic alignment module does not constrain candidate diversity during generation, but instead resolves the final output by favoring the candidate most compatible with the surrounding geographic context.

\section{Experiment}
\subsection{Dataset}

To support training and evaluation for satellite-to-soundscape generation, we constructed SatSound-Bench, a multimodal benchmark that aligns environmental audio recordings with high-resolution satellite imagery and textual scene descriptions. An overview of the benchmark is shown in Figure~\ref{fig:satsoundbench}.

Field recordings were collected in multiple cities across China, Sri Lanka, Thailand, Myanmar, Malaysia, Singapore, Indonesia, Cambodia, Vietnam, and the Philippines. Recordings were captured using vehicle-mounted setups, including a Zoom F6 multitrack recorder, external directional and omnidirectional microphones, and an Insta360 X4 camera with a USB-C microphone adapter. Audio was standardized to mono, resampled, and segmented into 10-second clips at 48\,kHz, yielding high-quality geotagged soundscape samples.

These field recordings were supplemented with environmental audio from three public datasets—SoundingEarth \cite{Heidler2022SoundingEarth}, iNaturalist Sounds \cite{Chasmai2024iNatSounds}, and Freesound \cite{freesound}—to expand coverage across diverse acoustic environments. Corresponding satellite imagery was retrieved from Google Maps, cropped into $512\times512$ patches, and temporally aligned with the recordings within a $\pm 3$-month window. Textual scene descriptions were obtained by manually annotating the field recordings and then expanded using a large language model (GPT-5.2), while captions for public datasets were generated directly by the model. Audio--text pairs were further filtered using CLAP similarity, retaining pairs with scores greater than 0.5 to ensure high-confidence matches.

In total, SatSound-Bench contains 28,630 satellite--text--audio pairs across 13 scene categories, with 24,400 pairs used for training and 4,230 pairs held out for testing. The benchmark supports evaluation protocols for assessing audio quality and geographic consistency.

\subsection{Implementation Details}

All experiments were implemented in PyTorch. Dense patch-level features were extracted from satellite imagery using a satellite-pretrained DINOv3 ~\cite{Simeoni2025DINOv3} vision transformer (v3-vitl16-sat493m). For structural geospatial attributes, dense DINOv3 patch embeddings were clustered with K-means ($k=8$). Each cluster was represented by 9 visual statistics concatenated with a 1024-dimensional DINO centroid, yielding a 1033-dimensional input to the random forest classifier. The two-stage random forest used 300 trees, a confidence threshold of 0.70, and a minimum cluster area ratio of 0.01. The final image-level geographic descriptor consisted of five dimensions: vegetation coverage, water ratio, built-up ratio, road density, and land-use mix.

Textual scene descriptions were used as inputs to eight text-to-audio generators: AudioLDM \cite{text2audio2}, AudioLDM2 \cite{text2audio3}, Auffusion \cite{xue2024auffusion}, Tango2 \cite{text2audio7}, Make-An-Audio~2 \cite{huang2023makeanaudio2}, EzAudio \cite{EzAudio2024}, AudioX \cite{Tian2025AudioX}, and MeanAudio \cite{Li2025MeanAudio}. Audio embeddings were extracted with the CLAP model \cite{audio_language1} using the laion/clap-htsat-unfused encoder. High-dimensional embeddings were projected into a 32-dimensional subspace using PCA for robustness.

The geo--acoustic alignment module was implemented as a two-layer MLP with a hidden dimension of 256 and GELU activations. During inference, each scene produced six candidate audio samples via the acoustic semantic expansion strategy, which were then scored and ranked using the geo--acoustic alignment module. All experiments were conducted on eight NVIDIA RTX Pro 6000 Blackwell GPUs (96GB).

All components are implemented using lightweight architectures to ensure scalability. In particular, the geographic descriptor is limited to five dimensions, and the projection network remains shallow, enabling efficient inference while maintaining alignment performance.

\begin{table*}[t]
\small
\centering
\setlength{\tabcolsep}{4.5pt}
\renewcommand{\arraystretch}{1.1}
\caption{Main comparison with recent image-to-audio and multimodal-to-audio baselines on the SatSoundBench test split. Results include both objective metrics and human evaluation.}
\label{tab:main_results}
\begin{tabular}{llccccccccc}
\toprule
\multirow{2}{*}{Generation Type} & \multirow{2}{*}{Method}
& \multicolumn{6}{c}{Objective Metrics}
& \multicolumn{3}{c}{Human Evaluation} \\
\cmidrule(lr){3-8}
\cmidrule(lr){9-11}
&
& FAD$\downarrow$ & FD$\downarrow$ & CLAP$\uparrow$
& KL$\downarrow$ & OVL$\uparrow$ & IS$\uparrow$
& MOS-A$\uparrow$ & MOS-S$\uparrow$ & MOS-E$\uparrow$ \\
\midrule

\multirow{4}{*}{Image-to-Audio}
& SSV2A \cite{guo2024ssv2a} 
& 7.53 & 46.96 & 0.214 & 0.622 & 0.617 & 2.570 
& 2.18 $\pm$ 0.71 & 2.05 $\pm$ 0.68 & 2.14 $\pm$ 0.73 \\
& Seeing and Hearing \cite{vision2audio2} 
& 11.32 & 51.26 & 0.233 & 0.633 & 0.569 & 2.747 
& 2.31 $\pm$ 0.82 & 2.22 $\pm$ 0.79 & 2.27 $\pm$ 0.76 \\
& IM2Wav \cite{Im2Wav2023} 
& 12.01 & 41.81 & 0.110 & 0.769 & 0.559 & 3.334 
& 2.04 $\pm$ 0.77 & 1.93 $\pm$ 0.74 & 2.01 $\pm$ 0.72 \\
& See-2-sound \cite{Dagli2024See2Sound} 
& 12.21 & 63.03 & 0.047 & 1.510 & 0.347 & 4.610 
& 2.42 $\pm$ 0.80 & 2.36 $\pm$ 0.83 & 2.48 $\pm$ 0.79 \\
\midrule

\multirow{3}{*}{Multimodal-to-Audio}
& CoDi \cite{tang2023codi2} 
& 14.02 & 59.09 & 0.021 & 1.249 & 0.425 & 3.241 
& 2.09 $\pm$ 0.75 & 1.98 $\pm$ 0.72 & 2.06 $\pm$ 0.74 \\
& AudioX \cite{Tian2025AudioX} 
& 13.10 & 46.42 & 0.082 & 0.773 & 0.536 & 3.810 
& 2.56 $\pm$ 0.98 & 2.43 $\pm$ 0.81 & 2.61 $\pm$ 0.77 \\
& AudioGenie \cite{rong2025audiogenie} 
& 3.53 & 18.43 & 0.435 & 0.185 & 0.815 & 2.410 
& 2.83 $\pm$ 0.72 & 2.69 $\pm$ 0.75 & 2.88 $\pm$ 0.70 \\
\midrule

& \textbf{Ours}
& \textbf{1.765} & \textbf{12.060} & \textbf{0.449}
& \textbf{0.098} & \textbf{0.847} & 2.480
& \textbf{3.58 $\pm$ 0.64} & \textbf{3.41 $\pm$ 0.67} & \textbf{3.66 $\pm$ 0.61} \\
\bottomrule
\end{tabular}
\end{table*}

\subsection{Evaluation Metrics}

We evaluate generated soundscapes using both objective metrics and human evaluations, as summarized in Figure~\ref{fig:pipeline}.

\textbf{Objective Metrics.}
We evaluate generated soundscapes using a set of objective metrics that capture distributional similarity, semantic alignment, and geographic consistency. Distributional similarity between generated and reference audio is measured using Fréchet Audio Distance (FAD) in the standard audio feature space and Fréchet Distance (FD) in the Pretrained Audio Neural Networks (PANNs) feature space~\cite{audio_representation1}, reflecting how closely the generated audio matches the reference audio in these embedding spaces. Semantic alignment between scene descriptions and generated audio is assessed using cosine similarity between text and audio embeddings extracted from CLAP~\cite{audio_language1}. Geographic consistency is evaluated using the GeoAlign score,
$\mathrm{GeoAlign}_i=\max_j \frac{a_{ij}^{\top} g_i}{\|a_{ij}\|_2 \, \|g_i\|_2}$,
where $a_{ij}$ denotes the embedding of the $j$-th candidate audio sample and $g_i$ denotes the projected geographic embedding of scene $i$.

To further characterize the generated audio distributions, we report Kullback–Leibler Divergence (KL), distribution overlap (OVL), and Inception Score (IS) based on predicted AudioSet class probabilities~\cite{gemmeke2017audioset}. Kullback–Leibler Divergence measures the difference between class probability distributions of generated and original audio, overlap quantifies the degree of agreement between distributions, and Inception Score captures semantic diversity and clarity of sound events.

\textbf{Human Evaluation.}
We recruited 20 participants to assess audio naturalness (MOS-A), sound--scene correspondence (MOS-S), and environmental immersion (MOS-E). For each participant, a stratified subset of 13 remote-sensing scene categories was randomly sampled, and generated clips from different methods were presented in random order under blind conditions. Ratings were provided on a five-point Likert scale, where 1 indicated very poor quality and 5 indicated excellent quality.

\subsection{Main Results}

We evaluate the proposed framework on the SatSoundBench benchmark constructed in this work. 
We first examine the effect of different text-to-audio backbones within our Satellite Imagery-to-Soundscape pipeline, and then compare the full framework with recent image-to-audio and multimodal-to-audio generation methods. All experiments are conducted on the test split.

\begin{figure*}[t]
  \centering
  \includegraphics[width=\textwidth]{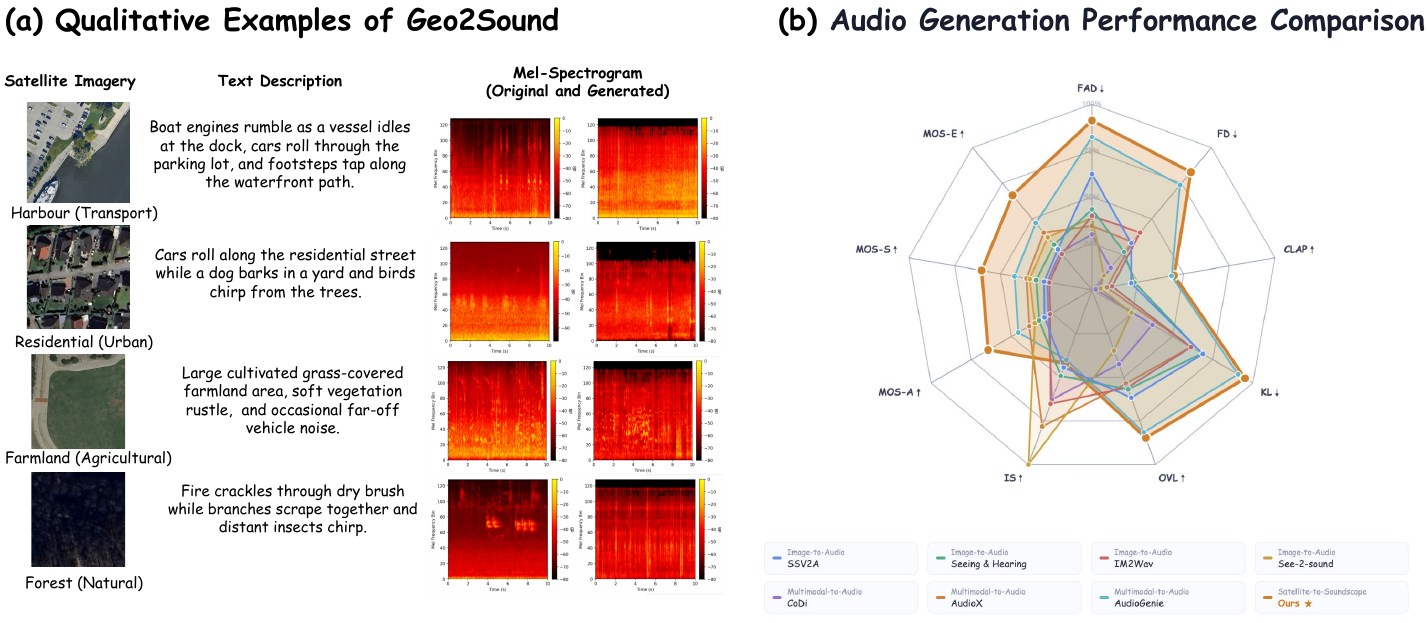}
  \caption{(a) Qualitative examples of Geo2Sound, showing paired satellite imagery, generated textual descriptions, and mel-spectrograms of original and synthesized audio, highlighting consistent scene-aware acoustic patterns across diverse environments. (b) Overall performance comparison across multiple objective and human evaluation metrics, demonstrating that Geo2Sound achieves balanced improvements in distributional similarity, semantic alignment, and perceptual quality.}
  \label{fig:mainresult}
\end{figure*}

\begin{table}[!ht]
\small
\centering
\setlength{\tabcolsep}{3pt}
\renewcommand{\arraystretch}{1.15}
\caption{Comparison of T2A backbones in the proposed satellite imagery-to-soundscape framework. All models used the same candidate textual inputs.}
\label{tab:t2a_backbones}
\resizebox{\columnwidth}{!}{
\begin{tabular}{lccccccc}
\toprule
Backbones & FAD$\downarrow$ & CLAP$\uparrow$ & FD$\downarrow$ & KL$\downarrow$ & IS$\uparrow$ & OVL$\uparrow$ & GeoAlign$\uparrow$ \\
\midrule
AudioLDM \cite{text2audio2} & 23.199 & 0.004 & 88.761 & 2.497 & 2.122 & 0.227 & 0.033 \\
AudioLDM2 \cite{text2audio3} & 2.663 & 0.360 & 14.314 & 0.113 & 2.654 & 0.840 & 0.360 \\
Auffusion \cite{xue2024auffusion} & 3.857 & 0.373 & 34.251 & 0.277 & 2.915 & 0.725 & 0.293 \\
Tango2 \cite{text2audio7} & 4.436 & \textbf{0.512} & 22.631 & 0.220 & 2.609 & 0.763 & \textbf{0.382} \\
Make-An-Audio 2 \cite{huang2023makeanaudio2} & \textbf{1.765} & 0.449 & \textbf{12.060} & \textbf{0.098} & 2.480 & \textbf{0.847} & 0.339 \\
EzAudio \cite{EzAudio2024} & 7.580 & 0.322 & 35.965 & 0.455 & \textbf{3.322} & 0.665 & 0.241 \\
AudioX \cite{Tian2025AudioX} & 5.733 & 0.417 & 26.586 & 0.306 & 3.123 & 0.722 & 0.310 \\
MeanAudio \cite{Li2025MeanAudio} & 4.65 & 0.451 & 14.63 & 0.210 & 2.20 & 0.751 & 0.290 \\
\bottomrule
\end{tabular}
}
\end{table}

\subsubsection{Effect of T2A Generation Backbones}

To analyze the role of the audio generator, we integrate several representative text-to-audio models into the same pipeline while keeping the upstream satellite imagery-to-text process fixed. In this way, the comparison mainly reflects the generation capability of each backbone under the same semantic inputs.

As shown in Table~\ref{tab:t2a_backbones}, different backbones exhibit clear trade-offs across metrics. Make-An-Audio~2 achieves the strongest overall performance, obtaining the lowest FAD (1.765), the lowest FD (12.060), the lowest KL divergence (0.098), and the highest OVL (0.847). These results indicate that its generated soundscapes are most similar to real environmental audio in both feature-space distribution and class-probability distribution. It also delivers strong semantic alignment, with a CLAP score of 0.449, which is among the best results across all evaluated backbones.

In contrast, Tango2 achieves the highest GeoAlign score (0.382), followed by AudioLDM2 (0.360), suggesting that these models produce candidates that are relatively more compatible with the geographic acoustic prior captured by our geo-acoustic alignment module. MeanAudio achieves the highest CLAP score (0.451), while EzAudio and AudioX obtain the highest IS (3.322 and 3.123, respectively), indicating that no single backbone dominates all metrics.

Overall, Make-An-Audio~2 provides the most balanced performance across distributional similarity, semantic alignment, and geographic consistency. We therefore adopt Make-An-Audio~2 as the default audio generation backbone in the subsequent experiments.

\subsubsection{Comparison with Audio Generation Baselines}

We further compare the proposed framework with recent image-to-audio and multimodal-to-audio generation methods, including SSV2A, Seeing and Hearing, IM2Wav, See-2-Sound, CoDi, AudioX, and AudioGenie. All methods are evaluated on the same test split of 4,230 satellite--audio pairs.

Table~\ref{tab:main_results} and Figure~\ref{fig:mainresult}(b) report both objective and subjective results. Compared with existing image-to-audio baselines, our method substantially improves distributional similarity and semantic alignment. In particular, it achieves the lowest FAD (1.765), the lowest FD (12.060), and the lowest KL divergence (0.098), while also obtaining the highest CLAP score (0.449) and the highest OVL score (0.847). These results indicate that the generated soundscapes are not only closer to real audio distributions, but also better aligned with the scene semantics inferred from satellite imagery.

Human evaluation further supports the objective results. Our method achieves the highest scores on all three Mean Opinion Score criteria, including MOS-A (3.58$\pm$0.64), MOS-S (3.41$\pm$0.67), and MOS-E (3.66$\pm$0.61). In comparison, the strongest baseline, AudioGenie, obtains 2.83$\pm$0.72, 2.69$\pm$0.75, and 2.88$\pm$0.70, respectively. This indicates that listeners consistently perceive our generated soundscapes as more natural, more scene-consistent, and more immersive.

Overall, the results demonstrate that combining acoustic semantic hypothesis expansion with geo-acoustic alignment substantially improves soundscape generation over existing image-to-audio and multimodal-to-audio baselines. Figure~\ref{fig:mainresult}(a) further presents qualitative examples, showing that the generated text descriptions and corresponding audio are consistent with the scene context across diverse environments. We next examine the contribution of individual components and the role of structural geospatial attributes inputs through a set of ablation studies.

\subsection{Ablation Study}

We conduct two complementary ablation studies to examine the contributions of structural geospatial attributes, the semantic expansion strategy, and the geo-acoustic alignment module in the proposed framework.

\begin{table}[t]
\small
\centering
\caption{Input ablation of structural geographic attributes for geo-to-acoustic embedding alignment. Higher validation cosine similarity and lower validation loss indicate stronger geographic--acoustic correspondence.}
\label{tab:projection_ablation}
\resizebox{0.9\columnwidth}{!}{%
\begin{tabular}{lcc}
\toprule
Setting & Best Val Cosine$\uparrow$ & Best Val Loss$\downarrow$ \\
\midrule
Main (full geo) & \textbf{0.324} & \textbf{0.676} \\
Single Road & 0.136 & 0.864 \\
Zero Input & 0.030 & 0.970 \\
Shuffled Geo & 0.025 & 0.975 \\
\bottomrule
\end{tabular}%
}
\end{table}

\begin{table*}[t]
\small
\centering
\setlength{\tabcolsep}{4.2pt}
\renewcommand{\arraystretch}{1.08}
\caption{Component ablation of the proposed framework. We analyze the contributions of semantic expansion and geo-acoustic alignment module.}
\label{tab:ablation_components}
\resizebox{0.95\textwidth}{!}{%
\begin{tabular}{lccrrrrr}
\toprule
Variant & Semantic Expansion & Geo-acoustic Alignment & CLAP$\uparrow$ & FAD$\downarrow$ & FD$\downarrow$ & KL$\downarrow$ & OVL$\uparrow$ \\
\midrule
Base                   & \xmark & \xmark & 0.3983 & 2.2270 & 17.9654 & 0.1733 & 0.7973 \\
w/o Semantic Expansion & \xmark & \cmark & 0.4232 & 2.1799 & 16.4729 & 0.1610 & 0.8014 \\
w/o GeoAlign           & \cmark & \xmark & 0.4135 & \textbf{1.7612} & 13.1779 & 0.1107 & 0.8393 \\
Full Model (ours)      & \cmark & \cmark & \textbf{0.4487} & 1.7653 & \textbf{12.0596} & \textbf{0.0977} & \textbf{0.8470} \\
\bottomrule
\end{tabular}
}
\end{table*}

We first evaluate whether structural geospatial attributes provide meaningful signals for geographic--acoustic correspondence. The model aligns structural geospatial attributes with the CLAP embedding of the reference audio from the same location under four settings: \textit{Main} (full input), \textit{Single Road} (road density only), \textit{Zero Input} (all-zero input), and \textit{Shuffled Geo} (randomly mismatched inputs and targets). As shown in Table~\ref{tab:projection_ablation}, the full input achieves the best validation cosine similarity (0.324) and the lowest validation loss (0.676), while the single-road setting shows only partial effectiveness. The similarly weak performance of the zero and shuffled settings indicates that the improvement is driven by meaningful geographic--acoustic correspondence rather than incidental statistical patterns.

We next conduct a component ablation on the full framework by removing semantic expansion and geo-acoustic alignment individually. As shown in Table~\ref{tab:ablation_components}, both components improve the final results. The full model performs best overall, achieving the highest CLAP score (0.4487), the lowest FD (12.0596), the lowest Kullback--Leibler divergence (0.0977), and the highest overlap score (0.8470). In contrast, removing semantic expansion leads to worse semantic alignment and weaker distributional similarity, indicating that multiple acoustic hypotheses help cover the diverse soundscape possibilities of the same satellite scene. Overall, semantic expansion and geo-acoustic alignment play complementary roles: the former increases soundscape diversity, while the latter helps select the most suitable output.

\section{Conclusion}

In this paper, we propose Geo2Sound, a scalable geo-aligned framework for satellite-to-soundscape generation. Unlike existing image-to-audio methods designed mainly for object-centric ground-level scenes, Geo2Sound addresses three key challenges of overhead imagery: the lack of structured geospatial semantics, one-to-many acoustic ambiguity, and the lack of broader geospatial context. To address these challenges, the framework combines structural geospatial attributes modeling, semantic hypothesis expansion, and geo-acoustic alignment in a unified generation framework.

To support this task, we further establish SatSound-Bench, the first large-scale benchmark of paired satellite imagery, text, and audio datasets, built from real-world field recordings and complementary public datasets. Extensive experiments and human evaluations show that Geo2Sound produces soundscapes that are more semantically plausible and geographically consistent than existing image-to-audio and multimodal-to-audio baselines. Overall, this work provides a first step toward geographically aligned soundscape generation from satellite imagery. 

This capability opens up novel avenues for urban planning and environmental management. By translating visual land-use patterns into auditory experiences, our framework can be integrated into digital twin cities ecosystems to simulate and evaluate the acoustic environment of large-scale urban zones. Crucially, this allows the public to intuitively experience the sonic consequences of issues like deforestation or urban sprawl, promoting deeper environmental engagement. 
In addition, this research holds potential for enhancing virtual reality and immersive media. By synthesizing comprehensive sensory experiences of remote or inaccessible locations, it promises to revolutionize domains such as virtual tourism and environmental education.

\begin{acks}
The project is supported by the National Natural Science Foundation of China (No. 42501268) and the Guangdong Natural Science Foundation (No. 2026A1515012525).
\end{acks}

\bibliographystyle{ACM-Reference-Format}
\bibliography{sample-base}

\appendix
\setcounter{table}{0}
\setcounter{figure}{0}
\renewcommand{\thetable}{S\arabic{table}}
\renewcommand{\thefigure}{S\arabic{figure}}

\section{Prompt Design and Hypothesis Analysis}

To analyze the effect of semantic hypothesis expansion, we compare the exact prompt settings used in our experiments with a controlled rephrasing variant. Let $C_0$ denote the base caption generated from the satellite image by GPT-5.2. This caption is shared across all settings. Our method further expands $C_0$ into two acoustically distinct hypotheses, denoted by $C_1$ and $C_2$, while the control setting generates two acoustically invariant rephrasings, denoted by $C_1'$ and $C_2'$.

Table~\ref{tab:prompt_analysis} compares three settings: \emph{Basic Caption}, which uses only $C_0$; \emph{Control}, which uses acoustically invariant rephrasings; and \emph{Ours}, which uses acoustically distinct yet visually plausible hypotheses. Both \emph{Control} and \emph{Ours} improve over the \emph{Basic Caption} baseline, indicating that prompt expansion is beneficial overall. Meanwhile, \emph{Ours} achieves slightly stronger alignment-oriented performance, especially on GeoAlign and CLAP, suggesting that the main benefit comes from constructing acoustically distinct hypotheses rather than merely changing the wording.

The exact prompt templates are shown in Table~\ref{tab:prompt_templates}. Our prompt preserves visual consistency while explicitly requiring differences in the dominant acoustic condition, whereas the control prompt keeps the dominant acoustic condition unchanged and only allows wording-level variation.

\begin{table*}[t]
\small
\centering
\caption{Prompt templates used for semantic hypothesis expansion and control.}
\label{tab:prompt_templates}
\fbox{%
\begin{minipage}{0.96\textwidth}

\textbf{Ours Prompt.}

\textbf{Input:} Base caption $C_0$ generated from the satellite image.\\
\textbf{Prompt:}\\
Given the following satellite image caption:\\
\texttt{"\{C0\}"}\\

Generate TWO alternative sentences describing the same scene,\\
but reflecting DIFFERENT plausible acoustic conditions.\\

\textbf{Requirements:}
\begin{itemize}
    \item Each sentence must remain consistent with the original visual scene.
    \item Each sentence must be a single concise English sentence.
    \item Do NOT include prefaces such as ``This scene'', ``The image shows'', etc. Start directly with the main content.
    \item The TWO sentences must differ in the dominant acoustic condition in a way that is acoustically meaningful (e.g., busy vs.\ quiet traffic, with vs.\ without natural ambience, higher vs.\ lower overall ambient noise).
    \item Each sentence MUST contain at least ONE explicit acoustic cue (e.g., a traffic-intensity cue such as ``busy traffic'' vs.\ ``nearly traffic-free'', or a natural-sound cue such as ``birds chirping'' vs.\ ``no birds'' / ``wind ambience'', or an ambient-noise cue such as ``noticeably loud'' vs.\ ``very quiet'').
    \item Do NOT introduce unrealistic or clearly unsupported sound sources.
    \item If the caption does not strongly support traffic, prefer natural ambience differences instead.
\end{itemize}

\textbf{Output format (exactly two lines):}\\
(1) [Sentence]\\
(2) [Sentence]

\vspace{0.8em}
\hrule
\vspace{0.8em}

\textbf{Control Prompt.}

\textbf{Input:} Base caption $C_0$ generated from the satellite image.\\
\textbf{Prompt:}\\
Given the following satellite image caption:\\
\texttt{"\{C0\}"}\\

Generate TWO alternative sentences describing the same scene,\\
but keep the dominant acoustic condition EXACTLY the same.\\

\textbf{Requirements:}
\begin{itemize}
    \item Each sentence must remain consistent with the original visual scene.
    \item Each sentence must be a single concise English sentence.
    \item Do NOT include prefaces such as ``This scene'', ``The image shows'', etc.
    \item The dominant acoustic condition (sound source type + presence/absence + relative intensity) MUST be the same in both sentences.
    \item Do NOT change any acoustic-relevant words or phrases, including:
    \begin{itemize}
        \item traffic intensity words (e.g., busy/light/quiet, continuous vehicles) and any implied change in vehicle noise;
        \item presence/absence of natural sound sources (e.g., birds, wind, water ambience) or their intensity;
        \item overall ambient noise level qualifiers (e.g., loud/very quiet/noisy/low-noise).
    \end{itemize}
    \item Only change acoustic-invariant aspects, such as phrasing, syntactic structure, or non-acoustic visual details that do NOT imply a different sound source or intensity.
\end{itemize}

\textbf{Output format (exactly two lines):}\\
(1) [Sentence]\\
(2) [Sentence]

\end{minipage}
}
\end{table*}

\begin{table*}[t]
\small
\centering
\caption{Quantitative comparison of the basic caption, control expansion, and our semantic hypothesis expansion. Our method shows clear gains over the basic caption while remaining overall close to the control setting.}
\label{tab:prompt_analysis}
\setlength{\tabcolsep}{7pt}
\renewcommand{\arraystretch}{1.15}
\resizebox{0.95\textwidth}{!}{%
\begin{tabular}{lccccccc}
\toprule
Method & GeoAlign$\uparrow$ & CLAP$\uparrow$ & FAD$\downarrow$ & FD$\downarrow$ & KL$\downarrow$ & OVL$\uparrow$ & IS$\uparrow$ \\
\midrule
Basic Caption & 0.2829 & 0.4232 & 2.1799 & 16.4729 & 0.1610 & 0.8014 & \textbf{2.6558} \\
Control  & 0.3308 & 0.4405 & \textbf{1.7095} & 12.1782 & \textbf{0.0920} & \textbf{0.8495} & 2.4855 \\
Ours     & \textbf{0.3390} & \textbf{0.4487} & 1.7653 & \textbf{12.0596} & 0.0977 & 0.8470 & 2.4804 \\
\midrule
Ours vs. Basic Caption & +0.0561 & +0.0255 & +0.4146 & +4.4133 & +0.0633 & +0.0456 & -0.1754 \\
Ours vs. Control  & +0.0082 & +0.0082 & -0.0558 & +0.1186 & -0.0057 & -0.0025 & -0.0051 \\
\bottomrule
\end{tabular}
}
\end{table*}

\section{Geo Projection Training Details}
\label{sec:geo_mlp_details}

This section provides additional implementation details for the geo-to-acoustic projection model used in candidate selection. The geo-acoustic alignment module maps the scene-level geographic descriptor to the target audio embedding space used for candidate selection. 
Following the main paper, the input is a 5-dimensional geographic descriptor, and the target is a CLAP-based audio embedding projected into a 32-dimensional PCA space. 
The projection network is implemented as a two-hidden-layer MLP with architecture $5 \rightarrow 256 \rightarrow 256 \rightarrow 32$, using GELU activations and dropout with rate 0.1.

We train the projection with AdamW, using a learning rate of $1\times10^{-3}$ and weight decay of $1\times10^{-4}$. 
The batch size is 64, and training runs for up to 80 epochs with early stopping (patience = 12). 
Early stopping is determined based on validation loss. 
A fixed random seed is used for Python, NumPy, PyTorch, and CUDA, affecting model initialization and training stochasticity.

Audio targets are extracted using the CLAP HTSAT-unfused encoder from 10-second audio clips at 48 kHz. The original CLAP embeddings are projected into a 32-dimensional PCA space, where PCA is fitted on the training split only.
All geographic input features are z-score normalized using statistics computed from the training split, and the same normalization is applied during validation and inference.

The output embedding of the MLP is L2-normalized, and the model is trained with a cosine-based regression loss between the predicted embedding and the target embedding in the PCA space. The cosine loss is computed on normalized embeddings.

We use a 15\% validation split with random seed 42. 
At inference, candidate audios are ranked by cosine similarity between the projected geographic embedding and the corresponding candidate audio embeddings, and the highest-scoring candidate is selected as the final output. 
For semantic-expansion settings, the candidate set consists of six audios generated from three caption hypotheses (the base caption $C_0$ and two expanded hypotheses), with two samples generated for each hypothesis.

\begin{table*}[t]
\small
\centering
\caption{Sensitivity analysis with respect to the number of candidate audios $N$. Increasing the candidate set improves performance initially, but the gains gradually saturate while inference time continues to grow. We therefore use $N=6$ in the main paper as a practical trade-off between performance and efficiency.}
\label{tab:candidate_sensitivity}
\setlength{\tabcolsep}{7pt}
\renewcommand{\arraystretch}{1.12}
\begin{tabular*}{\textwidth}{@{\extracolsep{\fill}}ccccccccc@{}}
\toprule
$N$ & GeoAlign$\uparrow$ & CLAP$\uparrow$ & FAD$\downarrow$ & FD (PANNs)$\downarrow$ & KL (PANNs)$\downarrow$ & IS (PANNs)$\uparrow$ & OVL (PANNs)$\uparrow$ & Time (min)$\downarrow$ \\
\midrule
1  & 0.1542 & 0.3953 & 2.2455 & 18.4886 & 0.1892 & \textbf{2.7417} & 0.7878 & \textbf{18.64} \\
3  & 0.2438 & 0.4127 & \textbf{2.0954} & 16.9107 & \textbf{0.1591} & 2.7187 & \textbf{0.8038} & 30.62 \\
6  & 0.2829 & 0.4232 & 2.1799 & 16.4729 & 0.1610 & 2.6558 & 0.8014 & 47.52 \\
10 & \textbf{0.3050} & \textbf{0.4292} & 2.1845 & \textbf{16.3595} & 0.1662 & 2.6319 & 0.8003 & 71.18 \\
\bottomrule
\end{tabular*}
\end{table*}

\section{Candidate Number Sensitivity}
\label{sec:candidate_sensitivity}

This section examines how the number of candidate audios affects alignment performance, distributional quality, and inference cost. Table~\ref{tab:candidate_sensitivity} analyzes the effect of the number of candidate audios $N$ used before geo-acoustic selection, where $N$ denotes the total number of generated candidate audios per scene. As $N$ increases from 1 to 6, the alignment-oriented metrics improve substantially, indicating that a larger candidate pool provides more opportunities for selecting geographically and semantically compatible outputs. When increasing $N$ further to 10, the gains become more marginal and several quality-related metrics no longer improve consistently, while inference time increases substantially. We therefore use $N=6$ in the main paper as a practical trade-off between performance and efficiency.

\begin{figure*}[t]
    \centering
    \includegraphics[width=\textwidth]{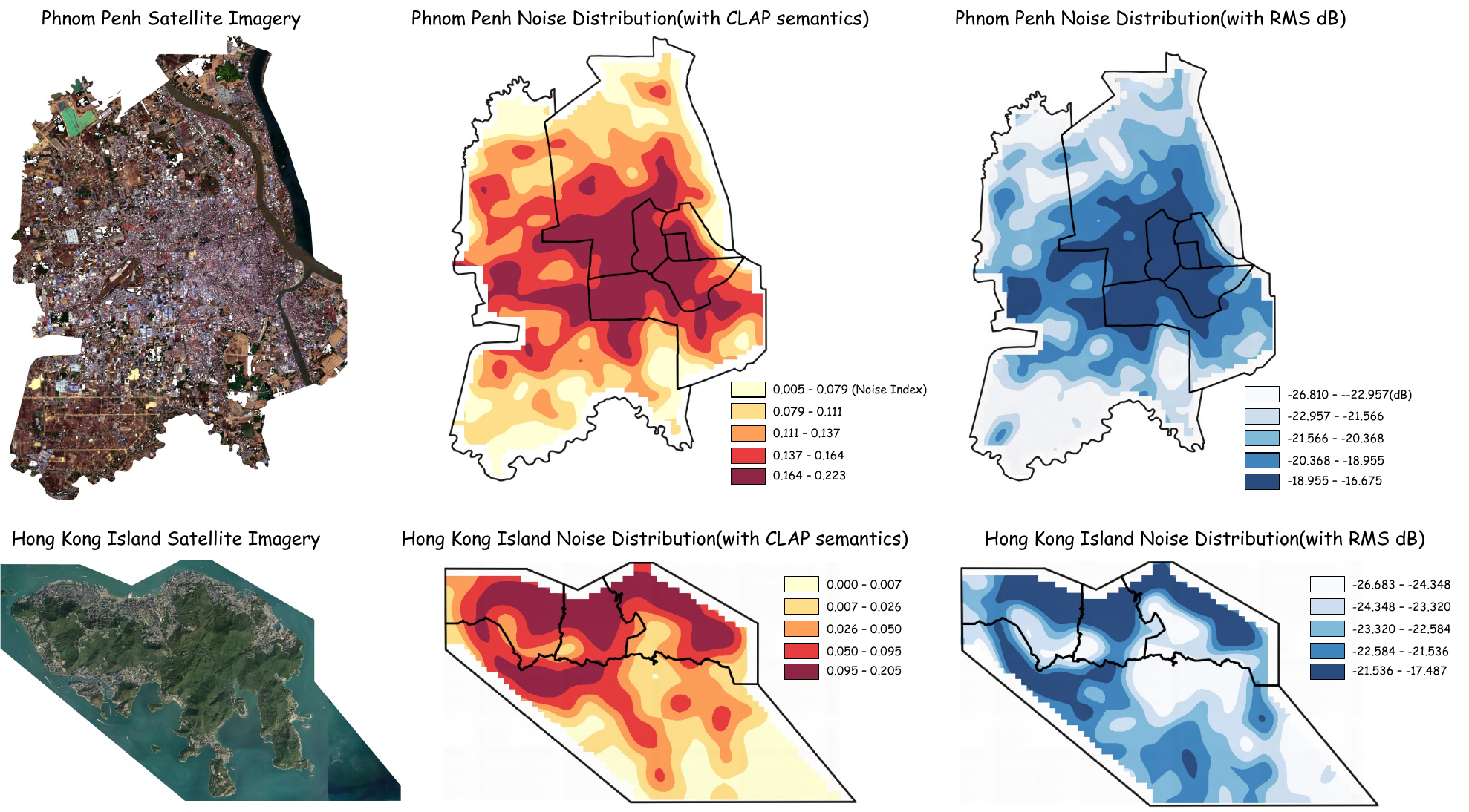}
    \caption{Scalable soundscape case studies in Phnom Penh and Hong Kong Island. Satellite imagery is shown together with noise distributions derived from CLAP semantics and RMS dB.}
    \label{fig:appendix_urban_case}
\end{figure*}

\section{Additional Qualitative Analysis}

\begin{figure*}[t]
    \centering
    \includegraphics[width=0.96\textwidth]{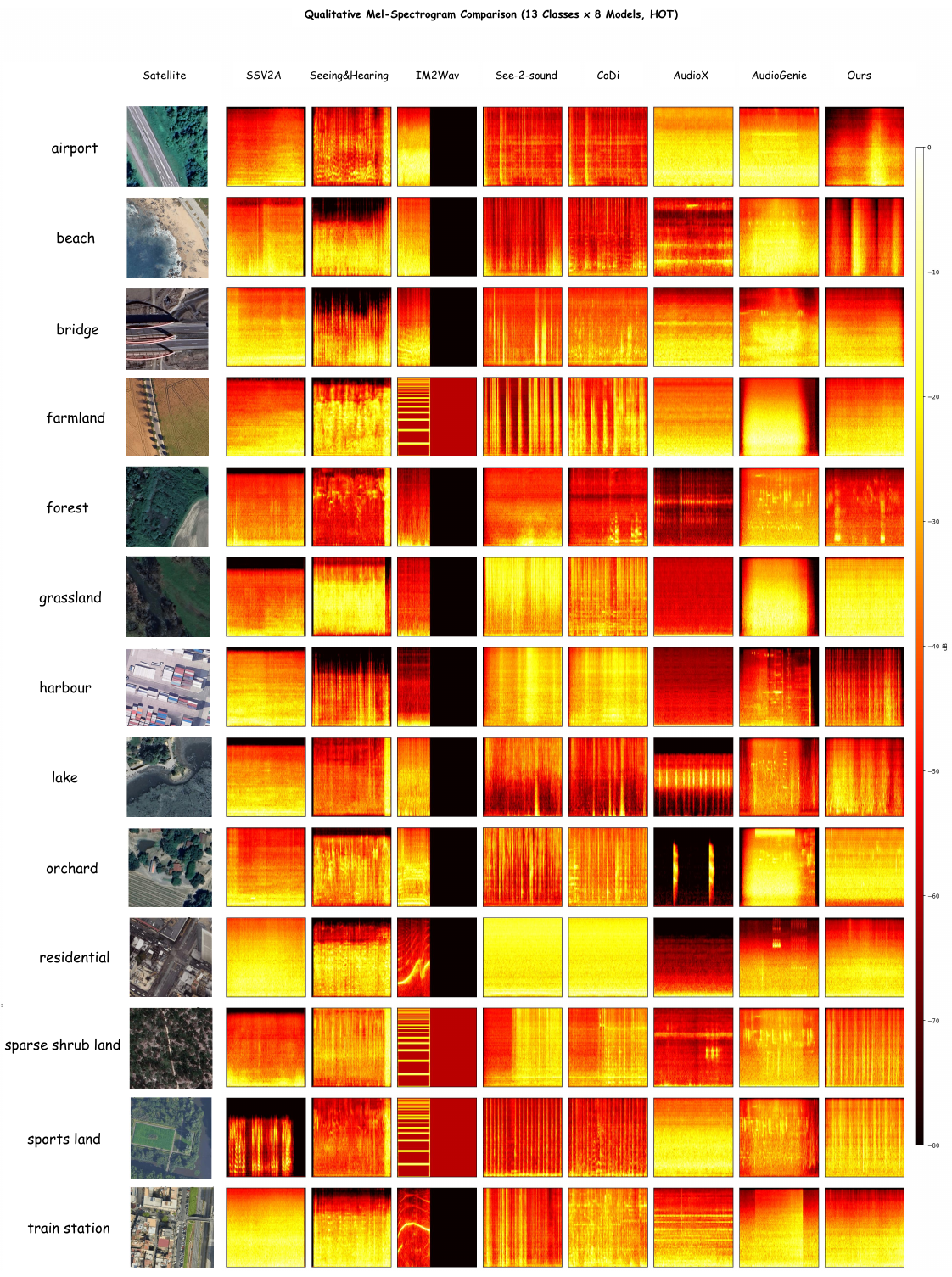}
    \caption{Qualitative mel-spectrogram comparison across 13 scene categories and 8 generation models. The figure complements the quantitative results by visualizing time--frequency differences across methods.}
    \label{fig:appendix_mel_comparison}
\end{figure*}

We provide additional qualitative comparisons to complement the main experiments across scene categories and large-scaled case studies. Figure~\ref{fig:appendix_mel_comparison} presents mel-spectrogram comparisons across 13 scene categories and 8 generation models, offering a time--frequency view of the generated soundscapes. Figure~\ref{fig:appendix_urban_case} further presents large-scale case studies in Phnom Penh and Hong Kong Island, illustrating spatial patterns in the generated soundscape distributions and highlighting the potential of our framework for scalable soundscape generation over large areas.

\begin{figure*}[t]
    \centering
    \includegraphics[width=\textwidth]{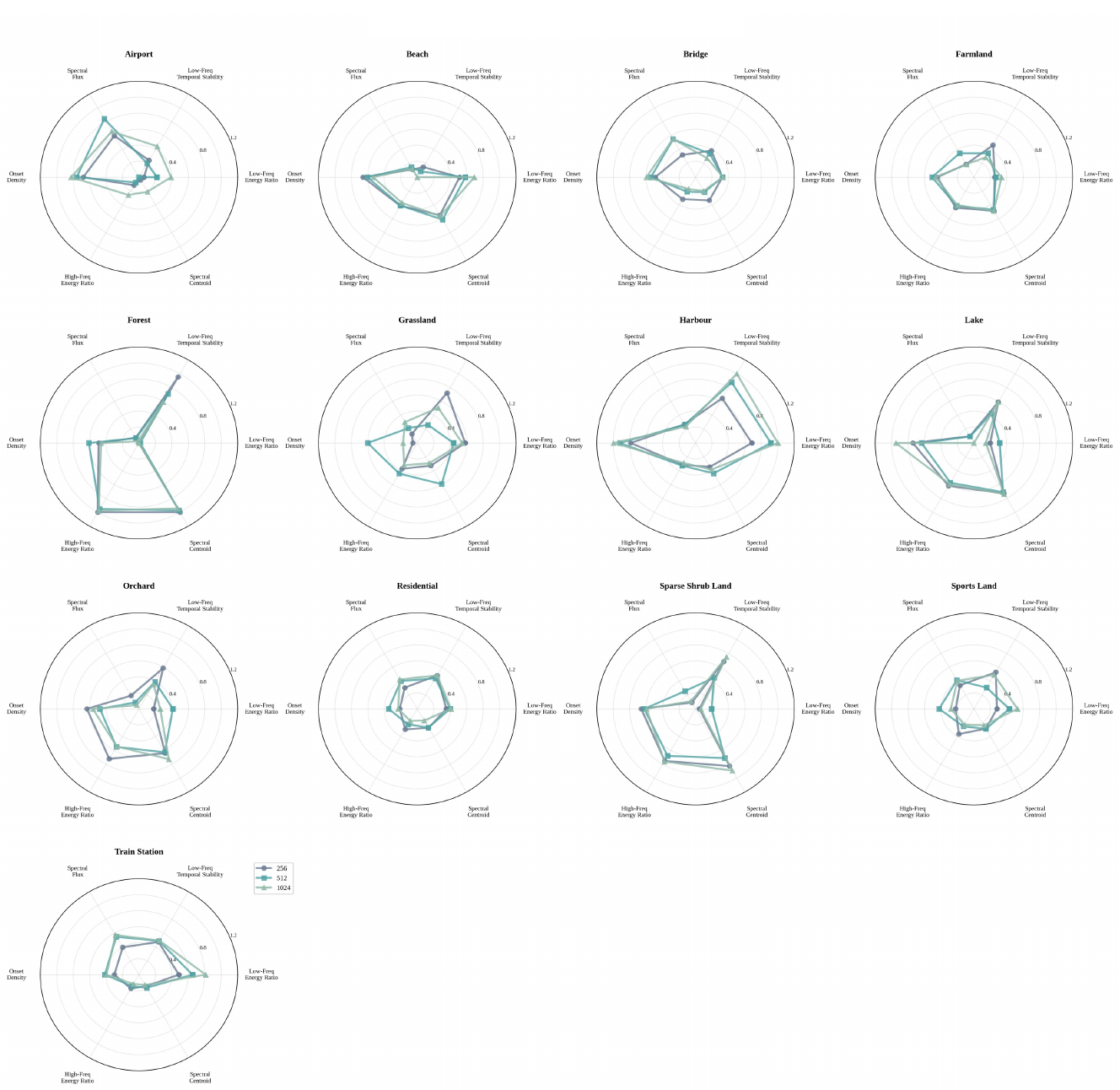}
    \caption{Supplementary scale-sensitive acoustic feature analysis across 13 scene categories under three image scales (256, 512, and 1024).}
    \label{fig:appendix_scale_radar}
\end{figure*}

\begin{figure*}[t]
    \centering
    \includegraphics[width=\textwidth]{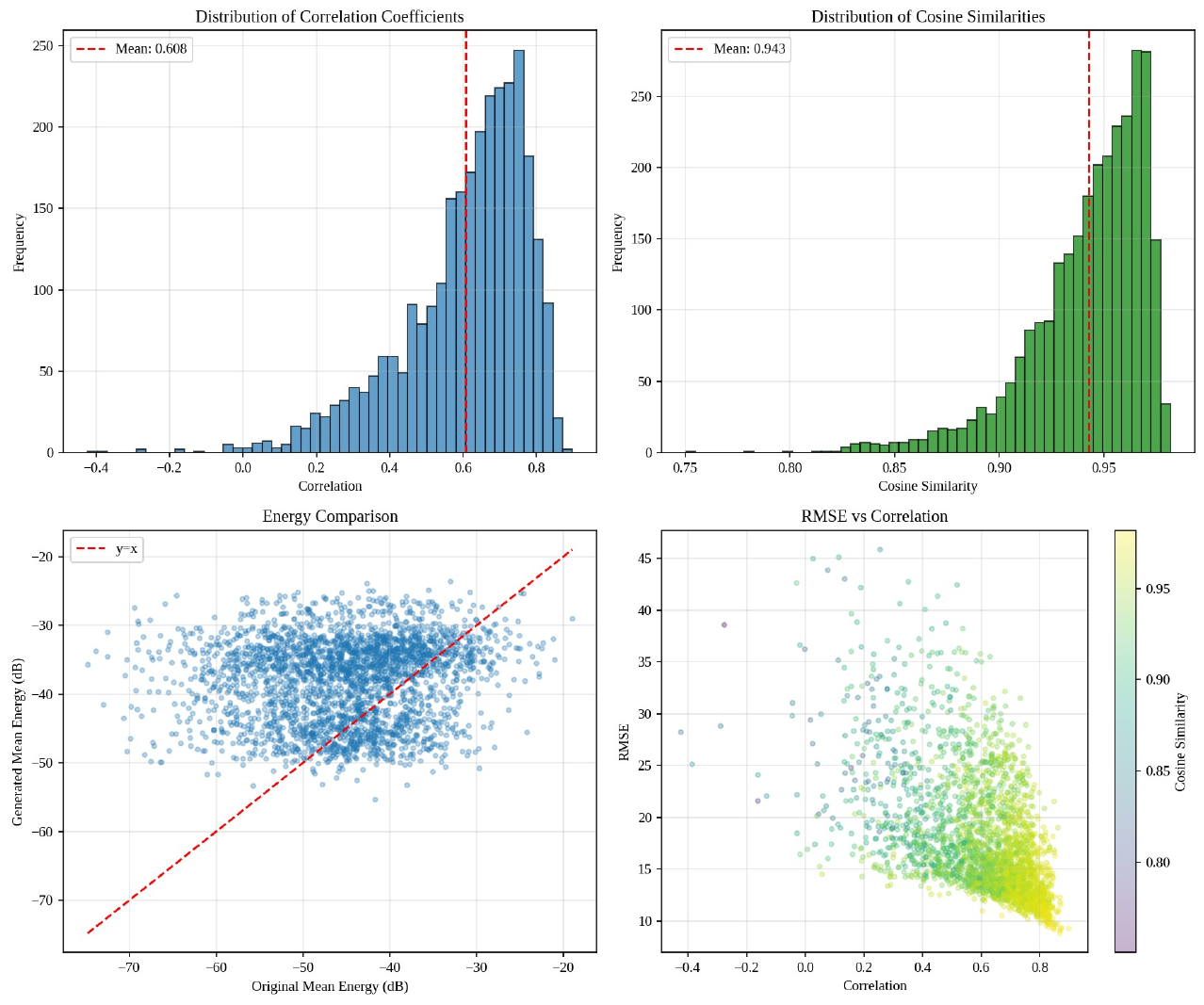}
    \caption{Supplementary global statistical analysis of generated and reference audio, including correlation, cosine similarity, energy, and RMSE relationships.}
    \label{fig:appendix_statistical_analysis}
\end{figure*}

\section{Supplementary Statistical Analysis}

We report additional statistical analyses to further examine the stability and consistency of the generated acoustic patterns. Figure~\ref{fig:appendix_scale_radar} compares normalized acoustic feature profiles across image scales for different scene categories, showing that the overall acoustic patterns remain broadly stable across scales, with only moderate variation in several categories. By comparison, inter-category differences are more substantial, indicating that scene type plays a more dominant role than image scale in shaping the generated acoustic profiles.
Figure~\ref{fig:appendix_statistical_analysis} summarizes global statistical relationships between generated and reference audio across all samples. The distributions show that correlation coefficients are generally positive and cosine similarities are consistently high, indicating broad correspondence in the learned acoustic space. The scatter plots further show that higher correlation is typically associated with lower RMSE, while mean energy exhibits an overall trend of correspondence without strict one-to-one matching. These results provide supplementary evidence that the generated soundscapes preserve meaningful global statistical consistency with the reference audio.


\end{document}